\documentclass[fleqn,usenatbib]{mnras}

\usepackage{newtxtext,newtxmath}

\usepackage[T1]{fontenc}
\usepackage{anyfontsize}
\DeclareRobustCommand{\VAN}[3]{#2}
\let\VANthebibliography\thebibliography
\def\thebibliography{\DeclareRobustCommand{\VAN}[3]{##3}\VANthebibliography}

\usepackage{xcolor}

\usepackage{graphicx}	
\usepackage{amsmath}	
\usepackage{bm}
\usepackage{xcolor}     
\definecolor{textRed}{rgb}{0.0, 0.0, 0.0}
\newcommand{\newtext}[1]{\textcolor{textRed}{#1}}
\definecolor{textGreen}{rgb}{0.00, 0.0, 0.00}

\title[Marked correlation environments]{Marked statistics across the cosmic web: Environmental dependent clustering in modified gravity simulations}

\author[Armijo et al.]{
Joaquin Armijo,$^{1,2,3}$\thanks{E-mail: joaquin.armijo@ipmu.jp}
Lucas Da Costa$^{4,5}$
\\
$^{1}$Kavli Institute for the Physics and Mathematics of the Universe (WPI),\\
The University of Tokyo Institutes for Advanced Study (UTIAS), The University of Tokyo, Chiba 277-8583, Japan \\
$^{2}$Center for Data-Driven Discovery, Kavli IPMU (WPI), UTIAS,
The University of Tokyo, Chiba 277-8583, Japan\\
$^{3}$Department of Mathematical Physics, Institute of Physics, University of São Paulo, R. do Matão 1371, 05508-090, São Paulo, SP, Brazil\\
$^{4}$ILANCE, CNRS – University of Tokyo International Research Laboratory, Kashiwa, Chiba 277-8582, Japan\\
$^5$Ecole Polytechnique, IP Paris, F-91128 Palaiseau,
France
}

\date{Accepted XXX. Received YYY; in original form ZZZ}

\pubyear{2025}

\begin{document}
\label{firstpage}
\pagerange{\pageref{firstpage}--\pageref{lastpage}}
\maketitle

\begin{abstract}
We study environment-dependent clustering using the marked correlation function applied to Hu-Sawicki $f(R)$ modified gravity simulations. This gravity theory enriches the structure formation by enhancing gravity in a scale-dependent form. By employing a multi-scale cosmic structure finder algorithm, we define the cosmic environments divided in: nodes, filaments, walls and voids. We find a stronger impact of modified gravity in nodes and filaments, which together dominate the information content by more than a factor of four relative to other environments. Combining environmental information further enhances the expected signal-to-noise ratio for CMASS- and DESI-like mock samples, particularly in configurations including filaments. Overall, marked correlation functions that incorporate environmental structure increase the information content by about a factor of two compared to standard density-based marks applied to the full galaxy sample. These results demonstrate the importance of environmental information, especially from filaments, in \newtext{improving the sensitivity of galaxy clustering tests to deviations from GR in the $f(R)$ scenario.}

\end{abstract}




\section{Introduction}\label{sec:introduction}

The Universe is described with an increasing accelerated expansion by the standard Lambda-Cold Dark Matter (\(\Lambda\)CDM) cosmological model \citep{Riess98,Perlmutter99}. While many observations of the cosmic microwave background (CMB) set the initial conditions for the model \citep{Aghanim2020}, several recent datasets, including results from the Dark Energy Spectroscopic Instrument \citep[DESI;][]{DESI_DR2_Results}, have hinted at tensions that challenge its completeness. Notably, DESI measurements have suggested a preference for dynamical dark energy over a pure cosmological constant, \(\Lambda\), when looking at the Baryon Acoustic Oscillation and full shape of the large-scale structure (LSS) of the Universe \citep{DESI_DDE,DESI_DDE2}. This points to a deeper mystery: the true nature of the mechanism driving the accelerated expansion of the Universe.

One alternative to \(\Lambda \) is the possibility that general relativity (GR) breaks down on cosmological scales, motivating the study of modified gravity (MG) theories \citep[][for a theoretical motivation]{Clifton2012}. These models introduce additional degrees of freedom to the gravitational sector, which could account for the late-time cosmic acceleration without invoking a cosmological constant. However, theoretical and observational constraints significantly limit viable MG models. For example, the almost identical propagation speed of gravitational waves and light \citep{LIGO1,LIGO2}, for instance, rules out many models that predict otherwise \citep{SaksteinJain2017,Creminelli&Filippo2017,Balgacem2018,Baker2021a}. The models that remain, must include a screening mechanisms such as the chameleon screening \citep{Khoury2004} and Vainshtein radius \citep{BabichevDeffayet} to reconcile modifications to gravity with stringent Solar System and astrophysical tests. Among the surviving models, $f(R)$ gravity \citep{DeFelice2010,Sotiriou2010} is one of the most studied due to its tunable screening and well-understood phenomenology \citep{Applebly2007,He2014,DeLaCruz2016,MacDevette2025}.

Testing MG models is challenging because the modifications are subtle and typically manifest in the highly non-linear regime of structure formation, where analytical methods lose accuracy \citep{Koyama2016,Aviles2021}. In this regime, $N$-body simulations become essential to study the imprints of MG on the matter distribution. Simulations of $f(R)$ gravity show that the fifth force enhances structure formation in low-density (unscreened) environments, while physics inside high-density regions remain largely unchanged due to screening \citep{Brax2013,Arnold:2019,Howard01102020}. These changes alter the geometry of the cosmic web, yielding deeper voids and more massive halos compared to GR \citep{Li2012,Cai2015}. Yet, disentangling these effects observationally is difficult due to degeneracies with galaxy bias and baryonic physics \citep{Arnold2019,Ellewsen2018}.

To evaluate the effects imprinted by modifications of gravity, traditional summary statistics such as the two-point correlation function or power spectrum are limited, as they primarily capture information from a Gaussian field. To recover information from the nonlinear regime, higher-order moments of the matter field, which becomes highly non-Gaussian at late times due to gravitational collapse must be explored. As a result of the nonlinear mechanisms involved in $f(R)$ gravity, these effects are also modified \citep{Oyaizu08,Oyaizu08b,Schmidt09}, motivating the use of several higher-order statistics to extract the impact of modified gravity on the large-scale structure. These include three-point estimators such as the bispectrum \citep{GilMarin2011,Bose2018}, higher-order moments \citep{Peel2018}, and non-Gaussian probes such as Minkowski functionals \citep{Ling2015,Fang2017,Jiang2024}, peak and void statistics \citep{Cautun2018,Paillas2018,Davies2024}, the scattering transform \citep{Valogiannis2024}, and marked statistics \citep{White:2016,Armijo2018,Hernandez-Aguayo2019,Aviles2021}. Among these, the marked correlation function, $\mathcal{M}$, has emerged as a particularly promising tool to probe modified gravity, while retaining the simplicity of a two-point estimator. By weighting galaxy pairs according to local environmental properties such as density or tidal fields, $\mathcal{M}$ enhances clustering signals and captures environmental dependencies that are often key signatures of modified gravity models.

The impact of modified gravity in individual cosmic structures is a definitive observational signature to test these models. By employing marked statistics, we can isolate how the new degrees of freedom (e.g. the fifth force) alter clustering in nodes, filaments, walls and voids separately. \newtext{This environmental sensitivity makes marked correlation functions a powerful and complementary probe of modified gravity. For tests that require modelling the interplay between the complex scale dependent screening mechanism of these modified gravity flavours and the cosmic web structures formation and evolution, simulation-based inference methods are increasingly favoured \citep{Baker2021b, Heymans2018}}. We propose a environmental dependent study of clustering in $f(R)$ simulations, applying marked correlation function to galaxies considering individual cosmic structures. We have two main objectives: First, to discern if there is any particular environment, where marked statistics is more sensitive to these modified gravity features; and second, \newtext{to quantify the additional sensitivity to deviations from GR} gained by incorporating environment-dependent clustering information from individual structures. Current observational constraints on the $f(R)$ amplitude $|f_{R0}|$ arise primarily from high-mass halo abundance \citep{Schmidt09, Cataneo2015, Vogt2024} and weak lensing peaks \citep{Liu2016}, with stage-IV forecasts discussed in \citet{Davies2024}, motivating the search for complementary probes with enhanced environmental sensitivity.

This paper is structured as follows. In Section~\ref{sec2}, we review the \(f(R)\) model of modified gravity and its implementation in \(N\)-body simulations, including the chameleon screening mechanism. Section~\ref{sec3} describes the simulations used for both GR and MG, as well as the construction of halo catalogues, mock galaxy samples and the estimator for the marked correlation function. In Section~\ref{sec4}, we define the cosmic web environments and outline the methodology to study the environmental clustering of each structure. Results and forecasts for individual environments are presented in Section~\ref{sec5}. We conclude in Section~\ref{sec6} with a summary and outlook for future work.

\section{The \textit{f(R)} theory of gravity} \label{sec2}

In the standard $\Lambda$CDM model the accelerated expansion of the Universe at recent times is driven by the cosmological constant $\Lambda$. In contrast, the $f(R)$ theory of gravity \citep{Sotiriou2010} introduces new physics that arises from the additional degree of freedom appearing in the equations of motion for gravity (see for example \citealt{Li2007}) leading to the same expansion of the Universe than in the standard paradigm. Then, this theory can be understood as an extension to the standard GR model, which can be tested in cosmological scales.

$f(R)$ introduces a function $f$, of the Ricci scalar, $R$, in the Einstein-Hilbert action

 \begin{equation}
     S = \int d^4x \sqrt{-g} \left( \frac{1}{2\kappa^2}\left[R + f(R)\right] + \mathcal{L}_m \right),\label{chp2:eq1}
 \end{equation}
 
where $k^2 = 8\pi G$, $G$ is Newton's constant, $g$ is the determinant of the metric $g_{\mu\nu}$ and $\mathcal{L}_m$ is the Lagrangian density of matter. The addition of this extra term in Equation~\ref{chp2:eq1} leads to the modification of all the equations of GR, including the Einstein field equations

 \begin{equation}
     G_{\mu\nu} + f_RR_{\mu\nu} - g_{\mu\nu} \left[ \frac{1}{2}f - \nabla^2 f_R \right] - \nabla_{\mu}\nabla_{\nu}f = 8\kappa T_{\mu\nu}, \label{chp2:eq2}
 \end{equation}
 where $\nabla_{\mu}$ is the covariant derivative of the metric tensor, $f_R \equiv {\rm d}f(R)/{\rm d}R$ is a scalar and dynamical new degree of freedom arised from $f(R)$. To obtain the equations of motion for massive particles, we solve the trace of Eqn.~\ref{chp2:eq2} for a perturbation around the standard Friedmann-Lema\^itre-Robertson-Walker metric leading to two equations of motion. The modified Poisson equation:

 \begin{equation}
     \Vec{\nabla}^2 \Phi = \frac{16\pi G}{3}a^2[\rho_m - \bar{\rho}_m] + \frac{1}{6}a^2 \left[ R(f_R) - \bar{R} \right], \label{chp2:eq3}
 \end{equation}
 and the one for the new scalar field, $f_R$:

 \begin{equation}
     \Vec{\nabla}^2 f_R = -\frac{1}{3}a^2\left[ R(f_R) - \bar{R} + 8\pi G(\rho_m - \bar{\rho}_m) \right], \label{chp2:eq4}
 \end{equation}
 where $\rho_{\textrm{m}}$ is the matter density field, and an overbar indicates quantities ($\bar{\rho}_m$ and $\bar{R}$) defined as mean values for the background cosmology. By defining the Ricci scalar as a function of $f_R$ in both Eqns~\ref{chp2:eq3} and \ref{chp2:eq4}, these two equations can combined to obtain:

 \begin{equation}
     \Vec{\nabla}^2 \Phi = 4\pi G a^2 \left[ \rho_m - \bar{\rho}_m \right] - \frac{1}{2}\Vec{\nabla}^2 f_R \label{chp2:eq5},
 \end{equation}

 which defines a new equation of motion for massive particles coupled to the new scalar degree of freedom. This new term can understood as the potential $- 1/2 f_R$ of an extra force, the fifth force, mediated by the scalar field $f_R$, which is referred as the scalaron \citep{GANNOUJI2012}.

\subsection{The chameleon mechanism}

\begin{figure*}
    \hspace*{2cm}\includegraphics[width=0.6\linewidth]{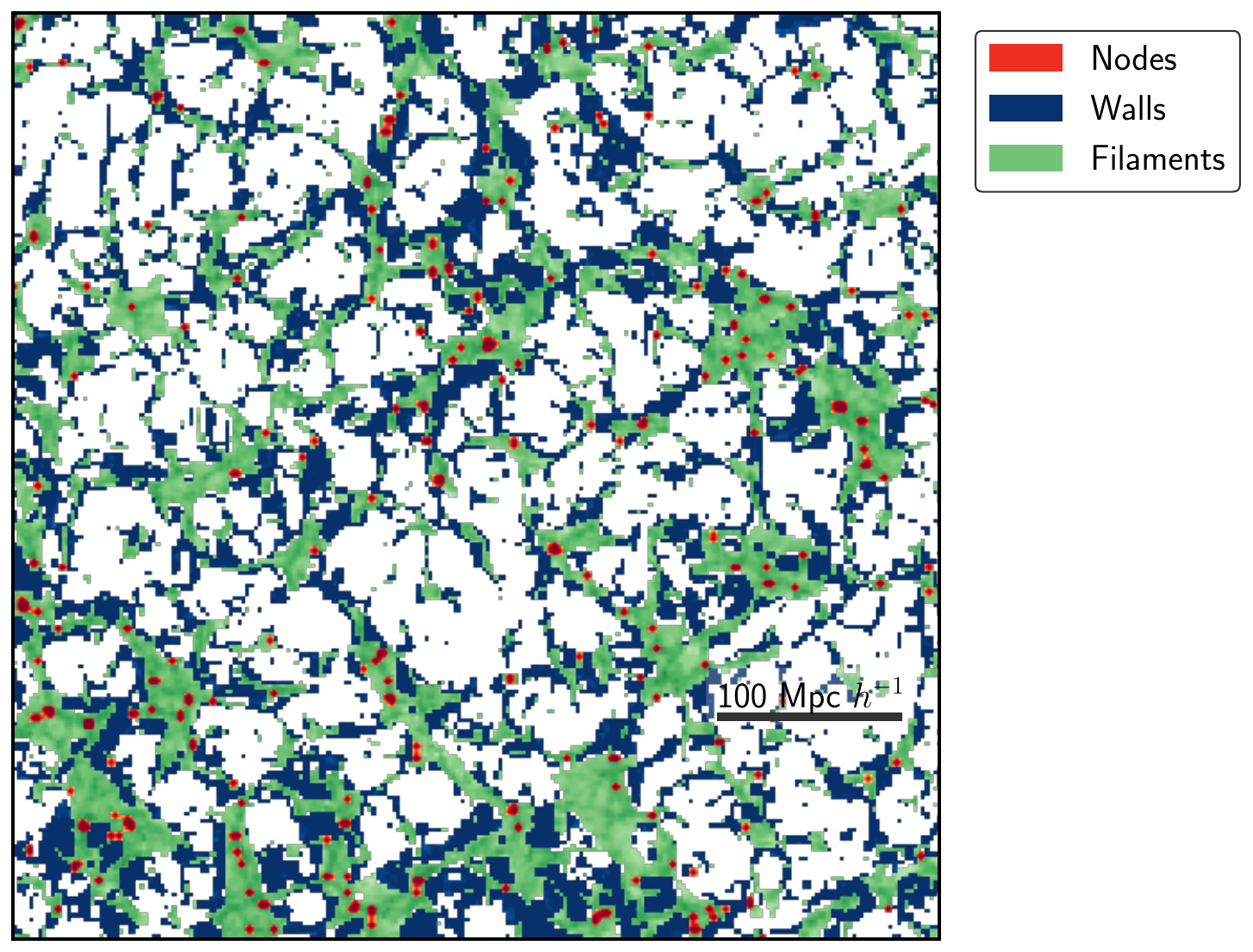} \\

    \caption{Cosmic web classification in a F5 modified gravity simulation. The cosmic structures are identified using the \texttt{pycosmommf} package based on the eigenvalues of the density field Hessian method. The visualization shows a slice of thickness $\Delta Z = 10 \, {\rm Mpc}\ h^{-1}$ from a simulation with box size of length $L=768\ {\rm Mpc}\, h^{-1}$. Structures are coloured as nodes (red), filaments (green), and walls (blue), while voids remain uncoloured (white).}
    \label{fig:SimEnvNexus}
\end{figure*}

To satisfy local tests where GR has been tested with high precision on certain scales—such as in the solar system \citep{Guo2014}, MG models must incorporate mechanisms that suppress the scalaron in Eqn.~\ref{chp2:eq5}. In $f(R)$ gravity, this suppression is achieved via a scale-dependent screening mechanism known as the chameleon mechanism \citep{Khoury2004}. On scales where the Newtonian potential is deep (e.g a massive galaxy or a large mass halo), Eqn.~\ref{chp2:eq4} is dynamically driven towards $\left| f_R \right| \rightarrow 0$. In this limit, Eqn.~\ref{chp2:eq5} is reduced to the standard Poisson equation, thereby recovering GR and ensuring the viability of the theory on such scales \citep{HuSawicki2007}. On the contrary, scales with a shallow Newtonian potential, the term $R - \bar{R}$ in Eqn.~\ref{chp2:eq4} becomes negligible, and Eqn.~\ref{chp2:eq5} simplifies to
\begin{equation}
    \Vec{\nabla}^2 \Phi = \frac{16}{3}\pi G a^2 \left[ \rho_m - \bar{\rho}_m \right],
\end{equation}

which corresponds to the Poisson equation enhanced by a factor of $4/3$, representing the unscreened, maximum amplitude of the fifth force, when the fifth force is present. Notably, no specific assumption about the functional form of $f(R)$ is required to obtain Eqn.~\ref{chp2:eq5}, making it independent of the choice of the $f(R)$ form.

 \subsection{The Hu \& Sawicki model}
 
A widely used choice for the functional form of $f(R)$ is the one proposed by \cite{HuSawicki2007} (HS, hereafter):
\begin{equation}
    f(R) = -m^2 \frac{c_1 \left( \frac{R}{m^2} \right)^n}{c_2 \left( \frac{R}{m^2} \right)^n + 1},
    \label{chp2:eq6}
\end{equation}
where $m^2 = 8\pi G \bar{\rho}_{m0} / 3$ is the mass scale, $\bar{\rho}_{m0}$ is the present-day background matter density, and $n$, $c_1$, and $c_2$ are free parameters of the model.  
This form is motivated by the requirement that, in the high-curvature regime ($R \gg m^2$), the ratio $m^2 / R$ becomes negligible, allowing $f(R)$ to be expanded as
\begin{equation}
    f(R) \approx -\frac{c_1}{c_2} m^2 + \frac{c_1}{c_2^2} m^2 \left( \frac{m^2}{R} \right)^n.
    \label{chp2:eq7}
\end{equation}
In the limit $m^2 / R \rightarrow 0$, the constant term $c_1 / c_2$ plays the role of an effective cosmological constant, independent of scale.  
Given this explicit form of $f(R)$, we can set $c_1 / c_2 = 6\Omega_{\Lambda,0} / \Omega_{m,0}$, where $\Omega_{m,0}$ is the present-day matter density parameter and $\Omega_{\Lambda,0} = 1 - \Omega_{m,0}$. With this choice, the model reproduces the $\Lambda$CDM background expansion history by construction.

The scalaron field is then approximated as
\begin{equation}
    f_R \approx -n \frac{c_1}{c_2^2} \left( \frac{m^2}{R} \right)^{n+1},
    \label{chp2:eq8}
\end{equation}
which can be evaluated today in the regime $R_0 \gg m^2$.  
In this case, the scalaron solution of Eqn.~\ref{chp2:eq4} sits at the minimum of the effective potential, and the Ricci scalar can be expressed using background values \citep{Brax2008}:
\begin{equation}
    \bar{R} \approx 8\pi G \rho - 2\bar{f}(R) = 3m^2\left[ a^{-3} + \frac{2}{3} \frac{c_1}{c_2} \right],
    \label{chp2:eq9}
\end{equation}
which removes the direct dependence between $R(f_R)$ and the scalaron $f_R$.  
This relation allows us to solve for $c_1 / c_2^2$ in Eqn.~\ref{chp2:eq8}:
\begin{equation}
    \frac{c_1}{c_2^2} = -\frac{1}{n} \left[ 3\left( 1 + 4\frac{\Omega_{\Lambda,0}}{\Omega_{m,0}} \right) \right]^{n+1} f_{R0},
\end{equation}
where $f_{R0}$ is the present-day value of the scalaron.  
With these choices, the model free parameters are two: $n$ and $f_{R0}$. These can be constrained using late-time large-scale structure observations. One of the key probes is the matter power spectrum, evaluated for models with varying $\left| f_{R0} \right|$ while keeping $n = 1$ fixed, providing the current constraints of this model \citep{Schmidt09}.

\begin{table*}
    \centering
    \begin{tabular}{c|ccccc||c|ccccc}
        GR   & $\log M_{\rm min}$ & $\log M_{0}$ & $\log M_{1}$ & $\sigma_{\log M}$ & $\alpha$ 
             & F5  & $\log M_{\rm min}$ & $\log M_{0}$ & $\log M_{1}$ & $\sigma_{\log M}$ & $\alpha$ \\
        \hline\hline
        HOD1 ($z=0.0$)  & 13.117 & 13.152 & 13.953 & 0.220 & 0.935 
              &     & 13.142 & 13.239 & 14.028 & 0.239 & 1.028 \\
        HOD2  ($z=1.0$)& 12.700 & 13.800 & 12.850 & 0.15 & 1.080 
              &      & 12.770 & 13.780 & 12.950 & 0.100 & 1.120
    \end{tabular}
    \caption{5-HOD parameter for GR and F5 simulations. Two samples are created for both simulation in a box of $L_{\rm box} = 1536\, {\rm Mpc}\, h^{-1}$: HOD1 a CMASS-like sample with $n_{\rm gal} = 3.5\times 10^{-4}\ {\rm Mpc^{-3}}\ h^{3}$, and HOD2 being a DESI-LRG sample with $n_{\rm gal} = 4.9\times 10^{-4} \, {\rm Mpc^{-3}}\ h^{3}$, both with the same two-point clustering.}
    \label{tab:HODtable}
\end{table*}

\begin{figure}
    \centering
    \includegraphics[width=0.99\linewidth]{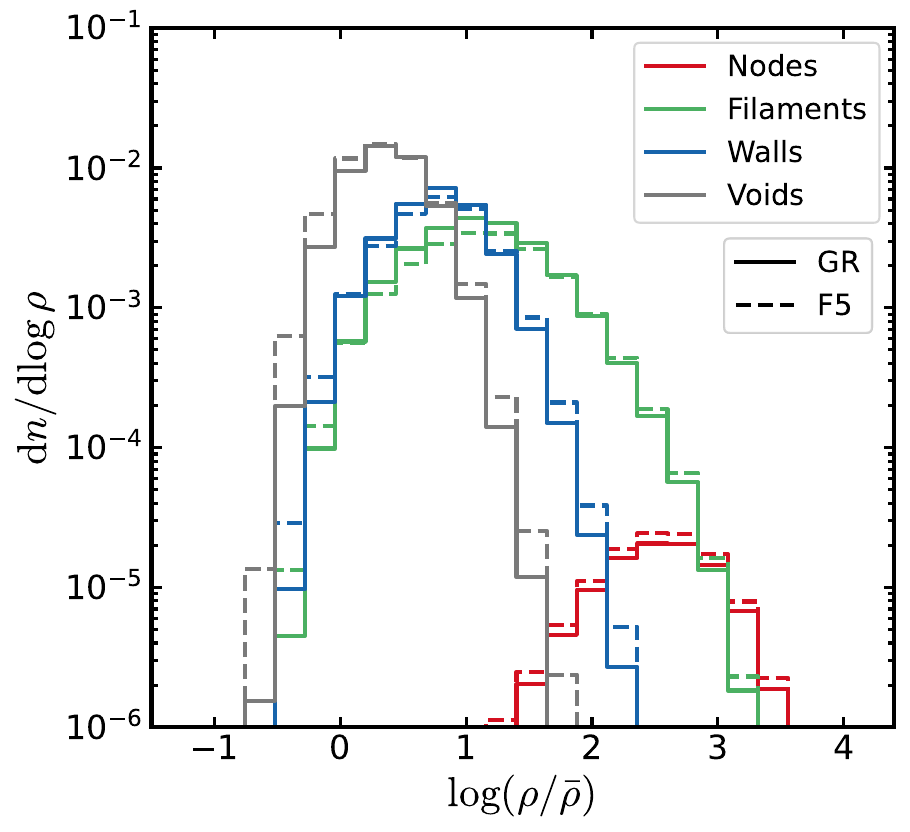}
    \caption{Histogram ${\rm d}n/d\log\rho$ of density values $\rho_i$ in both GR (solid lines) and F5 (dashed lines) simulations. We separate the $\rho_i$ values by their respective cosmic structure as defined by \texttt{pycosmommf}: Nodes (red), filaments (green), walls (blue), voids (grey). }
    \label{fig:dn_dlogrho}
\end{figure}

 \subsection{Marked correlation function}
 Originally implement as "Luminosity-weighted" correlation functions \citep{Skibba2006} have resurged due to their usefulness to test gravity and cosmology \citep{White:2016,Aviles2020}, using a function of the matter density contrast $\delta$ as weight or 'mark' $m$. It is defined as

 \begin{equation}
     \mathcal{M} = \frac{1}{n(r)\bar{m}^2} \sum_{ij} m_im_j, \label{eq:MCF1}
 \end{equation}
 with $n(r)$ the number of galaxy pairs at real-space separation $r$. In terms of of the standard two-point correlation function $\xi(r)$, $\mathcal{M}$ can be expressed as:

 \begin{equation}
     \mathcal{M} = \frac{1+W}{1+\xi},
 \end{equation}
where $W$ is the pair-weighted correlation function. In our case we focus on marks using an estimation of the local density $m = \rho^p$ and halo mass $m = M^p$, with $p$ a free parameter, which have been validated in several studies \citep{Armijo2018,Hernandez2018,Valogiannis2018}. For these marks, it has been shown that several definitions of $m$ as function of the matter field $\delta$ can be defined to enhance the information of either low-density and high-density regions, meaning it is an adequate test to probe the wide range of density values that define the environments of the cosmic web. Additionally, recent studies have found that $m$ can be optimized to extract the maximum amount of information, for a given definition of mark \citep{Karchner2025,Cowell2024}. Also, Marked correlation functions have been proven to constraint the cosmological parameters, with higher accuracy than the standard two point correlation functions in both real space \citep{Lai2024,Xiao2025} and power spectrum in Fourier-space \citep{Massara2023,Cowell2024}. 

For this study, our motivation is to find what is the role of the individual cosmic web environments in the modified gravity scenario, so we use the marks applied in \cite{Armijo2018}, which are defined to up-weight both low-density regions (e.g. cosmic voids) and high-densities (depending on the value of $p$), and at the same time focusing in environments that have been ignored before, such as filaments and walls. 

\section{Simulations.}\label{sec3}

We describe the the simulations used in this work, including the N-body dark matter particles used to estimate the density field, and the halo catalogues used to create the mock galaxy samples. 

\subsection{Simulations of modified gravity}
This analysis uses the modified gravity simulations of \cite{Arnold:2019}, which evolve $2048^3$ collisionless dark matter particles within a periodic cubic volume of side length $L_{\textrm{box}} = 1536\,h^{-1}\,\textrm{Mpc}$, corresponding to a particle mass of $M_{\textrm{p}} = 3.7 \times 10^{10}\,h^{-1}\,\textrm{M}_{\odot}$. We focus on the $z=0.0$ and $z=1.0$ outputs to directly assess the impact of modified gravity on the large-scale structure at two different times. The simulations adopt the 2016 \emph{Planck} cosmological parameters \citep{Planck:2016}: $h = 0.6774$, $\Omega_{\textrm{m}} = 0.3089$, $\Omega_{\Lambda} = 0.6911$, $\Omega_{\textrm{b}} = 0.0486$, and $\sigma_8 = 0.8159$. Two models are considered: an $f(R)$ run with scalaron amplitude $\left| f_{R0} \right| = 10^{-5}$, referred to as F5, and a standard General Relativity run, denoted GR. \newtext{This simulation suite was specifically designed for constructing light-cones that incorporate modified gravity effects \citep{Arnold:2019}, suited for detection-level forecast. A full constraint of fifth force parameter would require a suite designed for emulation of $f_{R0}$, such as \texttt{FORGE} \citep{Arnold2022} or \texttt{e-MANTIS} simulations \citep{SaezCesares2024}.}

\subsection{Density fields}

To estimate a density field that provides information about environment. We create a cloud-in-cell (CIC) density grid using \texttt{pylians} \citep{Pylians}. These densities are defined in a fixed comoving volume of $L_{\rm cell} = 2.19\, {\rm Mpc}\ h^{-1}$ resulting in a grid of $N_{\rm grid}=700\times 700\times 700$ densities, which will be used to define a density environment.

\subsection{Haloes}

To define haloes, we used \textsc{subfind} catalogues \citep{Springel:2001}. These are identified with a friends-of-friends (FoF) percolation scheme run on the fly for the simulation particles in a given snapshot. The minimum number of particles per group retained after the FoF step is set to 20 as used in \cite{Armijo2022}. Then, local density maxima are obtained from the FoF particle groups, conserving only the gravitationally bound structures and saved as subhaloes. Unbound particles are removed from the membership list. We save the positions of these haloes and subhaloes to populate galaxies when creating the mock catalogues reaching more realistic structures for the reconstructed mocks \citep{Armijo2024b}.

\begin{figure*}
    \centering
    \begin{tabular}{cc}
    \includegraphics[width=0.45\linewidth]{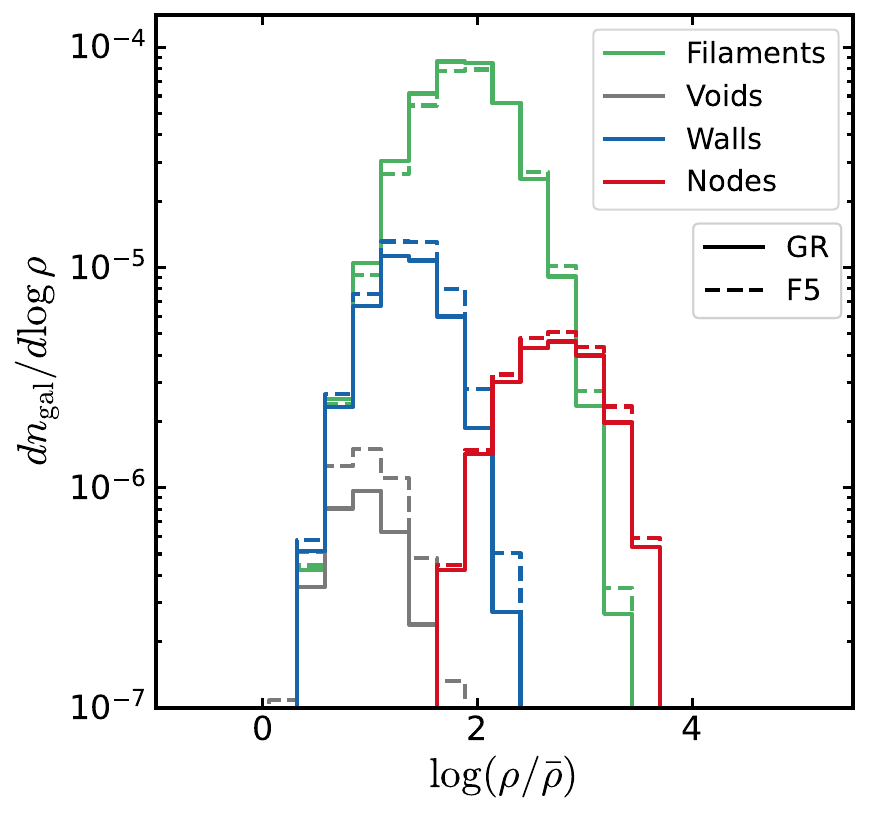}
    \includegraphics[width=0.45\linewidth]{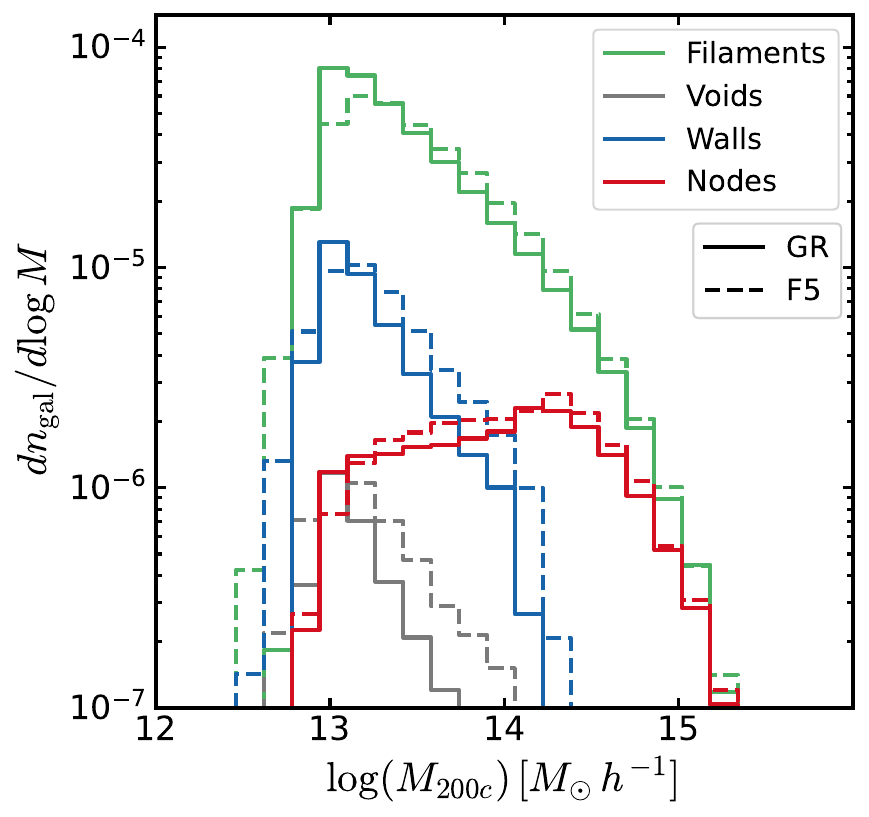}
    \end{tabular}    
    \caption{Left: Histogram of HOD galaxy density values ${\rm d}n_{\rm gal}/d\log\rho$ as function of $\log {\rho/\bar{\rho}}$ for the same cosmic structures of Figure \ref{fig:dn_dlogrho}. Right: Histogram of HOD galaxy mass values as function $\log M_{200c}$, the halo mass. Values are also divided by cosmic structure. These values used to mark galaxies to calculate the marked statistics.}
    \label{fig:galaxy-envs}
\end{figure*}

\subsection{HOD galaxy catalogues} \label{sec:SimHOD}

We use a 5-parameter HOD that has extensively used in modified gravity clustering studies \citep{Cautun2018,Armijo2018,Armijo2024b,Collier2024}. The HOD prescription \citep{Peacock2000,Berlind2002} is an empirical interpretation of the number of galaxies per halo as a function of halo mass. This is defined as \citep{Zheng2007}:

 \begin{eqnarray}
      \left< N_{\textrm{cen}} \right> & = & \frac{1}{2}\left[ 1 + \text{erf} \left( \frac{\log M - \log M_{\textrm{ min}}}{\sigma_{\log M}} \right) \right] \label{chp5:eq:HOD_cen} \\
     \left< N_{\textrm{sat}} \right> & = &\left< N_{\textrm{cen}} \right> \left( \frac{M-M_0}{M_1} \right)^{\alpha}. \label{chp5:eq:HOD_sat}
 \end{eqnarray}

 $N_{\textrm{cen}}$ is the mean number of central galaxies as a function of the mass of the halo, $M$, and $M_{\textrm{ min}}$ and $\sigma_{\log M}$ are free parameters. For satellites, the population of the halo is linked to whether or not there is a central galaxy. $M_0$, $M_1$, and $\alpha$ are free parameters. As presented in \cite{Armijo2024b} we use the position of sub halo catalogues provided by \texttt{subfind} to locate satellites distributed around the central galaxy, keeping the realism of the 1-halo term distribution for both simulations. We keep subhaloes up to scales we trust our simulations, which is around $r=0.5\, {\rm Mpc}\, h^{-1}$, below that we reach close to the simulation resolution limit of $\sim 30$ particles, which are removed as we only keep overdensities above $M_{\rm h} = 10^{12}\, M_{\odot} h^{-1}$. This limit is tested in \cite{Armijo2024a}, where an extensive study of the impact of the 1-halo in the two-point correlation function term is presented.

 The HOD parameters are selected using the posterior results obtained in \cite{Armijo2024b}, where model is selected to mimic an observational sample in both galaxy number density $n_{\rm gal}$ and real-space clustering between $0.5 < r /{\rm Mpc}\, h^{-1} < 80$. We calculate the two-point correlation function using the \texttt{Corrfunc} code \citep{Corrfunc2020}. These HOD catalogues are based on two different LRG samples: BOSS-CMASS and DESI LRG samples with $n_{\rm gal}^{\rm CMASS} = 3.5\times 10^{-4}\ {\rm Mpc^{-3}}\ h^{3}$ \citep{Anderson2012,Manera2013}, and $n_{\rm gal}^{\rm DESI} = 4.9\times 10^{-4} \, {\rm Mpc^{-3}}\ h^{3}$ \citep{Zhou2023,Zhou2023b}. The HOD parameter values are summarized in Table 1. \newtext{We acknowledge that the simulation snapshots used in this work at $z=0.0$ and $z=1.0$, do not precisely match the effective redshift of the observational samples they are intended to resemble: $z_{\rm eff} \approx 0.55$ for CMASS \citep{Anderson2012} and $z_{\rm eff} \approx 0.8$ -- $0.9$ for DESI LRGs \citep{Zhou2023}. However, we prefer to work directly with the outputs provided by the MG-lightcone project \citep{Arnold2019} for this type of comparison. Our primary motivation for adopting these redshift is twofold: first, to assess whether the modified gravity signal is preserved at higher redshift, where screening efficiency and halo properties differ; second, to construct galaxy mocks whose number densities are representative of current large spectroscopic surveys of LRG samples. Therefore we call CMASS-like and DESI-like mocks referred to number-density-matched analogues rather than precise reproductions of these surveys. And the results should be interpreted as indicative forecast accordingly}. We also consider the uncertainties found in \cite{Armijo2024a} related to considering all the valid HOD parameter combination when tuning them to replicate the two-point correlation function and number density of the observed galaxy samples, and fluctuations related to the random seed utilized to create the HOD catalogues. The former, can dominate the error-budget at the 1-halo term ($r < 1\ {\rm Mpc}\,h^{-1}$), as shown in \cite{Armijo2024a}, whereas they can also have an impact in the error bars of the marked correlation function. \newtext{Furthermore, to quantify the contribution of the HOD scatter to the marked correlation function, we compute 100 independent HOD realisations by varying the random seed used to populate haloes with galaxies, keeping the HOD parameters fixed. We emphasise that this test isolates the HOD stochastic contribution at fixed HOD parameters, and does not capture the full uncertainty associated with varying the HOD parameters themselves, which is explored separately in \cite{Armijo2024a}. Since these different runs assign galaxies to different haloes and subhaloes, these can contribute differently depending on their local density and host halo mass. We find that this contribution introduces uncertainties of $1$--$3\%$ for the marked correlation function on the scales studied in this work, which is subdominant with respect to estimated sample variance at such scales.}
 
\section{Definition of environment} \label{sec4}

\begin{figure*}
    \centering
    \begin{tabular}{c|c}
    \includegraphics[width=0.31\linewidth]{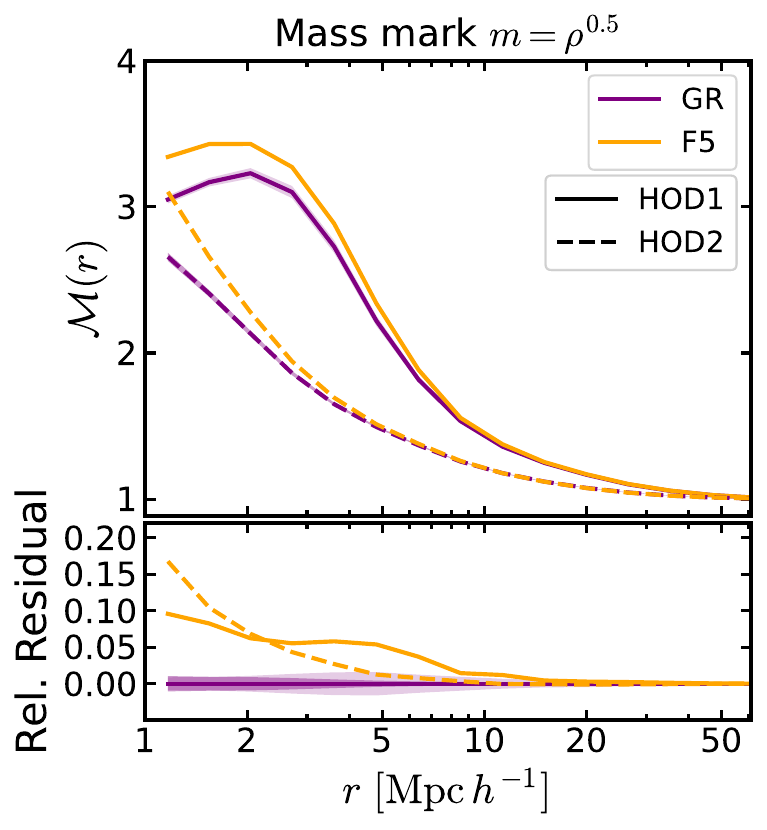}
        & 
    \includegraphics[width=0.64\linewidth]{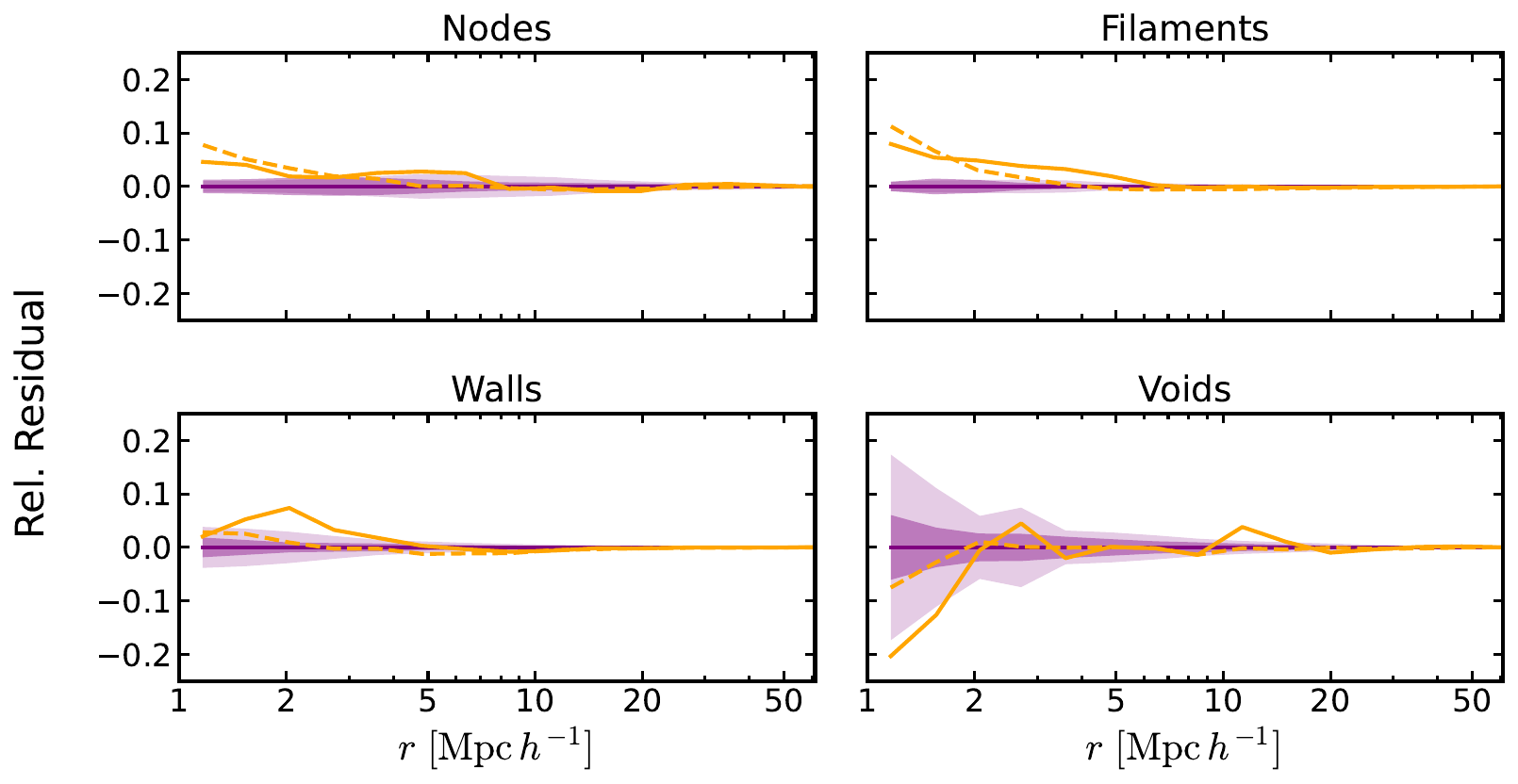} 
    \end{tabular}
    \caption{Left: Density-marked correlation function $\mathcal{M}$ as function of galaxy distance $r$ for GR (purple) and F5 (orange) simulations with $m=\rho^{0.5}$. We analyse results for HOD1 (solid) and HOD2 (dashed) samples in a range of distance between $1 < r / ({\rm Mpc}/h^{-1}) < 70$. We provide error bars for GR simulation as shaded region for HOD1 (light purple shade) and HOD2 (dark purple shade). A bottom subpanel showing the relative residual $(\mathcal{M} - \mathcal{M}^{\rm GR}) / \mathcal{M}^{\rm GR}$ to help visualisation. Right: Marked correlation function relative residuals for galaxies classified in different environments: Nodes (top-left), filaments (top-right), walls (bottom-left), and voids (bottom-right).}
    \label{fig:MarkedCF_density}
\end{figure*}

\newtext{The cosmic web classification used in this work is based on the eigenvalue analysis of the Hessian matrix of the smoothed density field, $\mathcal{H}_{ij} = \partial^2 \delta / \partial x_i \partial x_j$, evaluated at each point in the field. The eigenvalues $\lambda_i$ are then used to define to determine the local geometry of the region, with: $\lambda_1 < 0$ correspond to a sheet or wall, $\lambda_1 <0$; $\lambda_2 < 0$ for a filament, and $\lambda_1 <0$; $\lambda_2 < 0$; $\lambda_3 < 0$ for a cluster or node. Non-collapsing regions are classified as void structures. This approach provides a physically motivated decomposition of the large-scale structure into individual environments \citep{Hahn2007,ForeroRomero2009,Cautun2013}}. To define these cosmological environments, we use the the \texttt{pycosmommf} code for cosmic structure identification \citep{Sunseri2022}. This algorithm is based on the \texttt{NEXUS+} code \citep{Cautun2013}, which applies a multi-scale morphological filter to a smoothed density field for Hessian matrix computation. More details about \texttt{pycosmommf} can be found in \citep{Sunseri2022,Sunseri2025}. By sorting the eigenvalues of the Hessian matrix of the field $\rho$ at every location $\rho_i$, an environment is defined as part of a cosmological structure. These are nodes (all positive eigenvalues), filaments (2 positive eigenvalues), walls (1 positive) and voids (all negatives). We use a fix comoving size of $L_{\rm cell} = 2.19\, {\rm Mpc}\ h^{-1}$ to define a CIC density counting the number of massive particles from the simulation inside the cell, such as  \(\rho_{i} = N_i / L_{\rm cell}^3\), with $N_i$ the number of particles inside the $i$-cell. This value is selected to obtain a notion of environment in a non-linear scale similar to the definition of "small-scale" from \cite{Sunseri2025}. \newtext{We note that both the cosmic-web classification and the resulting marked correlation function may depend to some degree on this smoothing choice. A systematic exploration of the robustness of our conclusions with respect to the smoothing scale is beyond the scope of this work and is left for future analysis}. We show the classification obtained by \texttt{pycosmommf} in Figure~\ref{fig:SimEnvNexus}, where we compare a slice of the F5 simulation with $\Delta Z = 10.97\, {\rm Mpc}\ h^{-1}$. We paint the individual cells using this classification, to highlight the different structures. 
Once the cosmological structures are identified we will define a halo environment as the region where the halo lives with density $\rho_i$ and labelled by the respective cosmological structure. We also show the distribution of densities in Figure~\ref{fig:dn_dlogrho}, which compares the density values of GR and F5 simulations per structures. A small enhancement of low-$\rho$ values can be observed in the F5 simulation in filament, wall, and void environments, which is consistent with a fifth force enhancing the regions where the fifth force enhances gravity in MG models. All the other density values, for screened structures remain mostly unaffected.

\section{Environmental-dependent clustering using marked correlation functions}\label{sec5}

We use mock galaxy catalogues with the same $n_{\rm gal}$ and $\xi(r)$ as describe in Section \ref{sec3} to isolate the effect of MG in the studied environments. \newtext{We remind the reader that the CMASS-like (HOD1) and DESI-like (HOD2) labels refer to samples matched in number density to these surveys, analysed at $z=0.0$ and $z=1.0$ respectively}. Being the clustering mainly studied via the two-point correlation function, we match the clustering of our samples to find dependence in higher-order moments which are contained in the marked correlation function. We calculate the marked correlation function using the recipes from \cite{White:2016}, but with the definition of mark in the same fashion than \cite{Armijo2018,Hernandez2018}, highlighting two different marks: density dependent marks $m= (\rho/\bar{\rho})^p$ and host halo mass marks $m = M_{\rm h}^{p}$, we use the same values for the power index $p=0.5$ as they showed to correctly distinguish between modified gravity models in these previous studies. As we are not trying to optimize the mark, we keep these definitions and values throughout the whole paper. \newtext{The mark used throughout this work differs from the smoothed-density mark of \cite{White:2016} and \cite{Massara2023}, in which the local galaxy density field is explicitly smoothed with a kernel of size $R$ when defining the mark, introducing $R$ and $\delta_\star$ as free parameters. \cite{Satpathy2019} shows the choice of these parameters has little impact on the marked correlation signal. In the present work, we use the local density value inherited during the environment classification step, adopting a smoothing scale value of $R=8\ {\rm Mpc}\, h^{-1}$, which is representative of the range of scales used to characterise the cosmic web environments. This mark has been validated and used in previous work \citep{Armijo2024a}, and it is well suited to capture the environmental dependence of galaxy clustering in the context of modified gravity. While Satpathy et al. (2019) find that the marked signal is relatively insensitive to the smoothing scale, a dedicated assessment of how our environment-dependent results vary with this choice is left for future work.}

\subsection{Galaxy properties in environments.}

By creating mock catalogues that reproduce the clustering of CMASS (HOD1) and DESI (HOD2) LRG samples, we can directly test some of the density properties of these galaxies when considering the different environments where galaxies live. In Figure~\ref{fig:galaxy-envs}, we find a small excess of galaxies living in low-$\rho$ environments for F5 model in comparison to GR. This is expected as it has been previously shown that the formation of haloes is enhanced in MG models \citep{Cai2015}, which is consistent with these galaxies hosted in haloes living in voids. The same effect is also found in wall haloes with smaller enhancing. For galaxies in voids these also have higher masses as shown in the right panel of Figure~\ref{fig:galaxy-envs}, which is known to be an ideal probe to constraint modified gravity. Again, similar tendency is found in walls, filaments and the low-mass end of nodes. The latter represents low mass galaxies living in the outskirts of galaxy clusters which is also predicted to be enhanced, as it is currently used to constraint the amplitude of the fifth force using galaxy cluster abundance \cite{Schmidt09,Vogt2024,Cataneo2016,Liu2021} and weak lensing peak statistics \citep{Liu2016}.

\subsection{Environmental marked correlation functions.}

\begin{figure*}
    \centering
    \begin{tabular}{c|c}
    \includegraphics[width=0.31\linewidth]{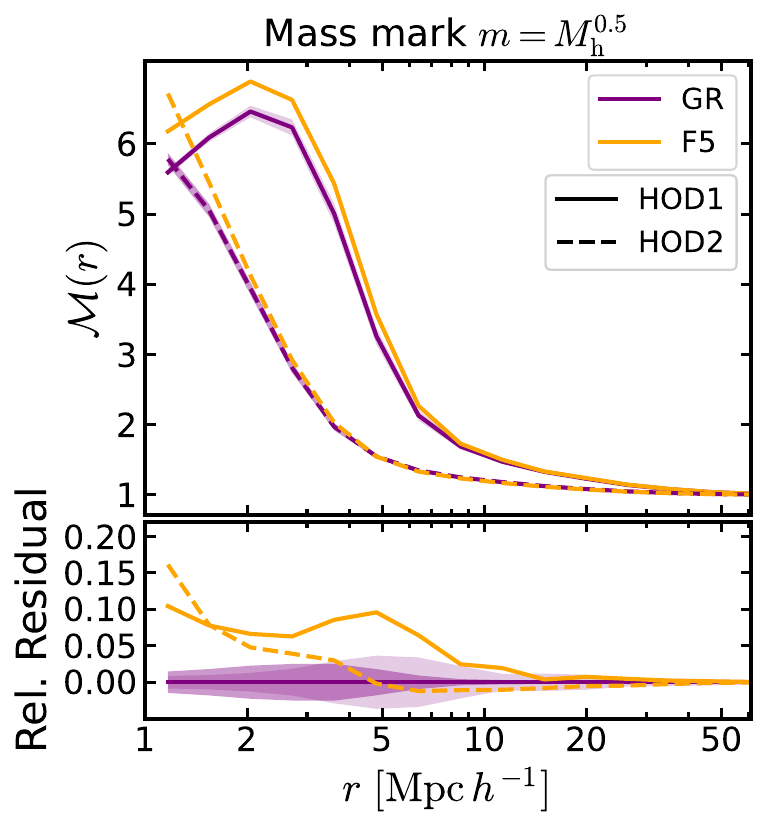}
        & 
    \includegraphics[width=0.64\linewidth]{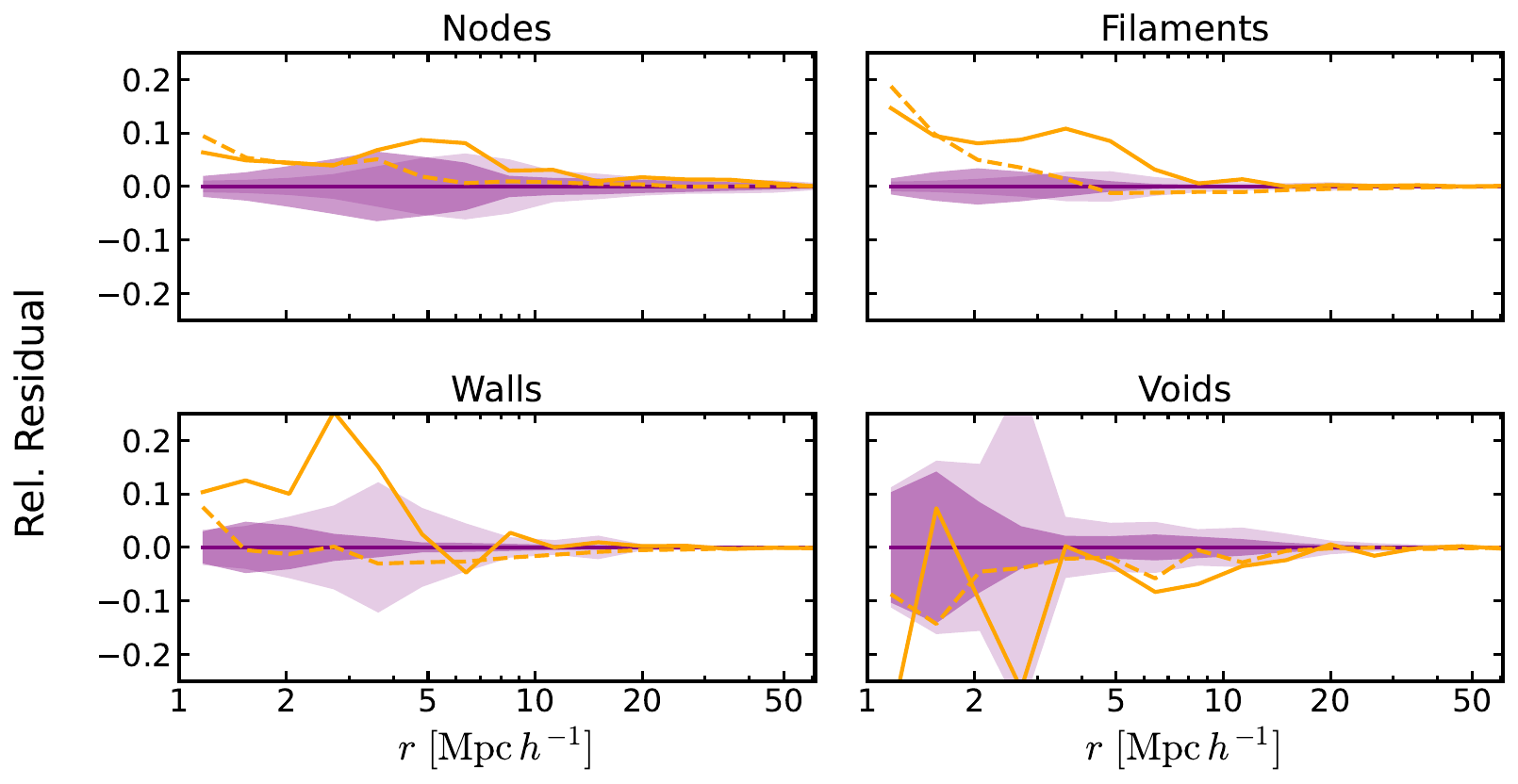} 
    \end{tabular}
    \caption{Same as Figure \ref{fig:MarkedCF_density} but for a mass-marked correlation function with $m=M^{0.5}$.}
    \label{fig:MarkedCF_mass}
\end{figure*}

We calculate marked correlation functions by dividing the galaxy samples in different environments, to discern if there is a preferred environment containing more information capable to distinguish modified gravity. We consider two scenarios: The fact that voids are predicted to be the most modified environment \citep{Clampitt_2012,Cai2015}, however with less galaxies. In the other hand, most of galaxies must live in filaments \citep{AragonCalvo2010}, which are more affected by MG than galaxies in nodes, either option must provide more information than galaxies in nodes. We also combine individual data vectors from marked correlation functions to see if the information content increases. 

We display results of density and mass marked correlation functions in Figure~\ref{fig:MarkedCF_density} and Figure~\ref{fig:MarkedCF_mass} respectively. First, we calculate $\mathcal{M}^{\rho}$ using density as a mark $m=\rho^{0.5}$, which upweights galaxies in overdensities. Additionally, we separate galaxies living in individual cosmic structures (right panels), to reinforce the environmental dependency of the marked correlation function. We estimate values for HOD1 and HOD2 mock samples that resemble CMASS and DESI LRG galaxies. In general, $\mathcal{M}^{\rho}$ is able to distinguish MG in most of environments (cosmic structures) and when using all galaxies (main plot, left-hand size). For both samples, deviations start from the smallest $r$-bin decreasing up to distances $r < 20\ {\rm Mpc}\,h^{-1}$. These deviations are stronger for a DESI-like sample (HOD2) at small scales, between $1 < r/({\rm Mpc}\, h^{-1}) < 2$. When analysing the marked clustering of individual cosmic structures, the tendency remains the same, specially for nodes and filaments where F5 simulation shows deviations up to 20\% from GR in a similar distance range. For walls and voids, $\mathcal{M}^{\rho,\,{\rm F5}}$ looks closer to GR simulations with a few data points showing differences between 5-10\%. Similar behaviour is found when using the mass mark $m=M^{0.5}$, but slightly changing the range where the deviations of MG are present to larger scales. This pattern is more prominent in walls a voids, where more differences can be identified. In particular, $\mathcal{M}^{M}$ for walls (in HOD2; dashed line) is around 5\% for $r>3\ {\rm Mpc} h^{-1}$, feature also present in $\mathcal{M}^{M}$ for voids, where F5 deviates from GR simulations in roughly 10\% between $5 < r/({\rm Mpc}\, h^{-1}) < 30$. However the size of the error bars (shaded area) also increases, particularly in voids. The fact that signature of MG gravity are found in all cosmic structures reveals the effectiveness of marked correlation functions as a environmental test at cosmological scale.
 
\subsection{Information content}

\begin{figure*}
    \centering
    \begin{tabular}{c|c}
    \includegraphics[width=0.45\linewidth]{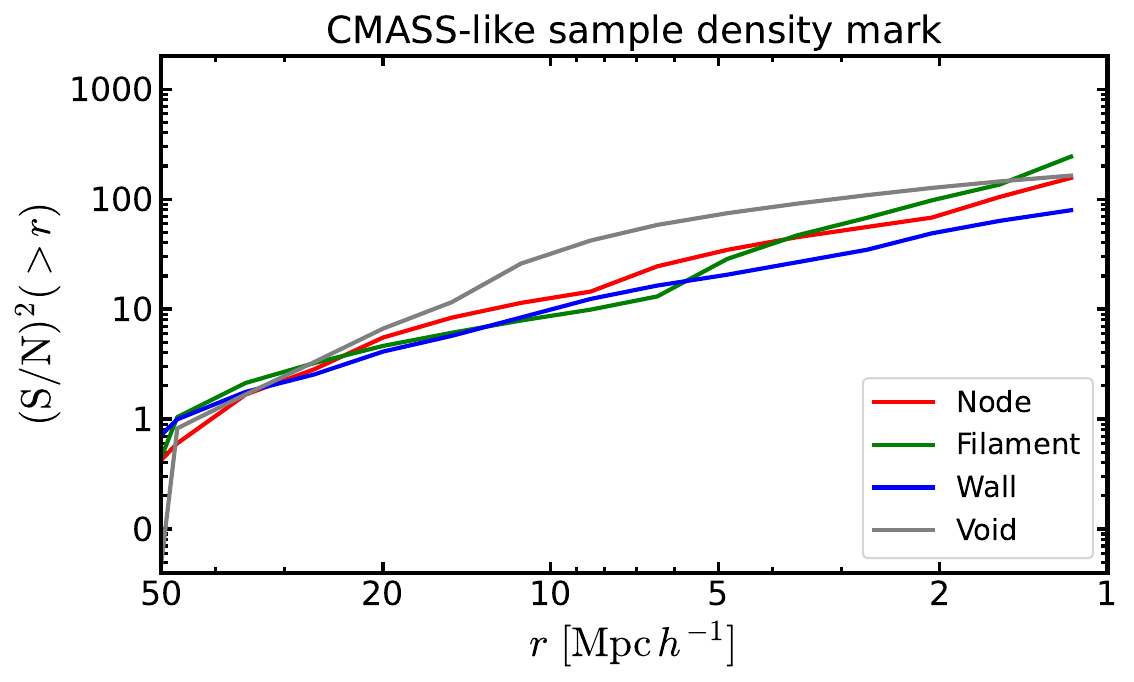}
         &
    \includegraphics[width=0.45\linewidth]{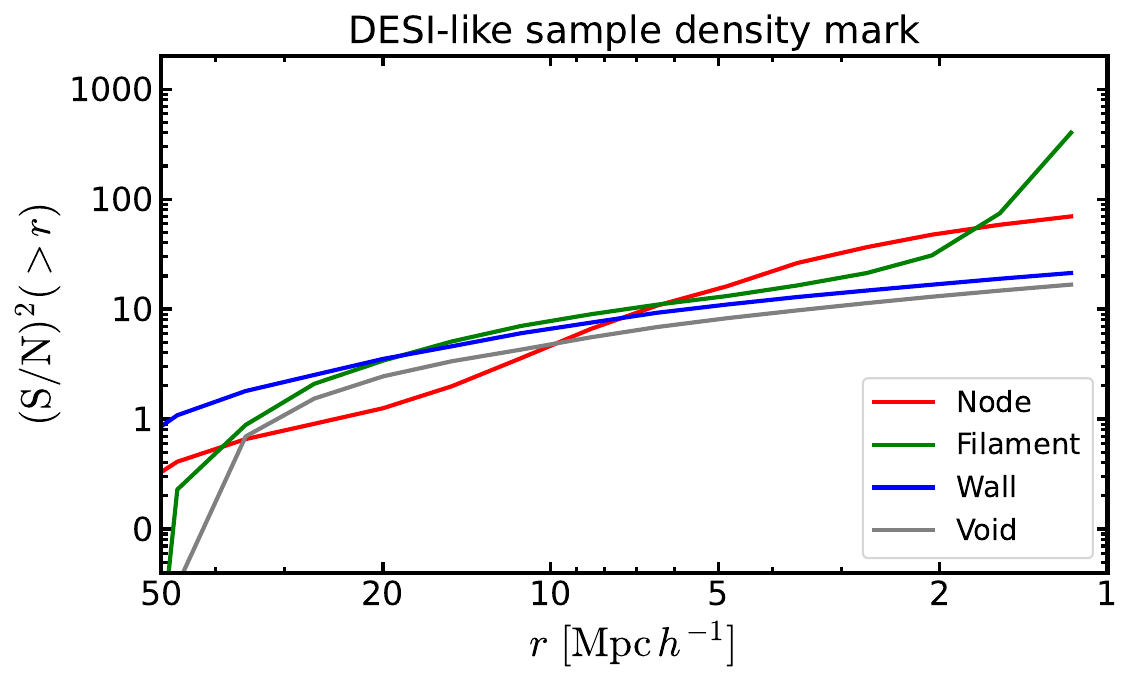}
    \end{tabular}
    \caption{signal-to-noise ratio as function of distance $r$ calculated using Equation \ref{eq:SNratio_distance}. We estimate these for different cosmic structures using same colour scheme than Figure \ref{fig:dn_dlogrho}. We sum elements starting from large $r$ to demarcate what scales are relevant for individual structures.}
    \label{fig:chisquareDistance}
\end{figure*}

To discern how powerful this test is, considering the individual cosmic structures and the surveyed galaxies, we need to apply a statistical tool that quantifies such significance. We choose to quantify deviations introduced in the F5 model as small modifications from the fiducial (GR) case by using the signal-to-noise (S/N) ratio test, also used in \cite{Cautun2018} (Eqn.~\ref{eq:SNratio_distance}), which defines a cumulative quantity as function of radial distance bin $r$:

\begin{equation}
    ({\rm S/N})^2 (> r) = \sum_{i\geq r; j\geq r}^{N_{\rm bins}}\, \delta \mathcal{M}_{i}^{\intercal} C_{ij}^{-1}\, \delta \mathcal{M}_{j} \label{eq:SNratio_distance}
\end{equation}

where $r$ is the separation bin where $\delta\mathcal{M}$ is integrated. However, differently from \cite{Cautun2018}, and by considering the nature of our observable $\mathcal{M}(r)$, which deviates at small $r$, we sum over from large $r$ values. \newtext{We use a covariance matrix $C_{ij}$ estimated using the Jackknife resampling method (see Appendix \ref{apx:Covariance} for a detailed calculation). We notice that this method can understimate uncertainties, particularly on scales where cosmic variance becomes important. However, we restrict the calculations to a range of scales where it is known that the two-point correlation function variance is similar for Jackknife, Bootstrap resampling, and Poisson realizations, $1.2 < r / \ ({\rm Mpc}\, h^{-1}) < 20$ \citep{Norberg2009}. For larger scales where sample variance might have an impact, the marked correlation function adds no additional information, as the values rapidly become unity. Because the covariance is estimated from a single simulation box via Jackknife resampling, the absolute values of $\chi_{\nu}^2$ and S/N reported below should not be interpreted as realistic survey forecasts. These errors derived from one realisation do not fully capture sample variance, and a robust forecast would require a covariance estimated from a large suite of independent mocks. We present these quantities primarily as relative comparisons between environments and between marks, for which the single-box estimate is sufficient to rank the constraining power of each configuration.} We plan to find two different things with this test: Find the particular scale where $\mathcal{M}$ becomes significant for different cosmic structures, and if any of the environments is particularly effective to find the deviations in MG simulations. 

We show the expected signal-to-noise as function of $r$ in Figure \ref{fig:chisquareDistance}, separating results for different cosmic structures, and estimating values for both HOD samples. For a CMASS-like sample (same number density and two-point function) all environments contain approximately the same information between $20 < r/({\rm Mpc}\, h^{-1})<50$. However, $r > 20\ {\rm Mpc}\, h^{-1}$, reveals a strong S/N ratio contribution from cosmic voids (grey line) in comparison to other structures. This is persistent up to small scales at $r>2\ {\rm Mpc}\, h^{-1}$, where filaments become more powerful (higher S/N values) and nodes have the same statistical information than voids. In the other hand, a DESI-like sample, presents a different behaviour for large scales, with nodes and voids clearly underperforming. However, both filaments and nodes become statistical significative in comparison to other structures at $r > 5\ {\rm Mpc}\, h^{-1}$. This clearly shows that filaments and nodes dominate the non-linear regime for both samples and redshift, whereas voids are relevant only at large (linear) scales at late times ($z=0.0$).

\begin{figure*}
    \centering
    \begin{tabular}{c|c}
    \includegraphics[width=0.445\linewidth]{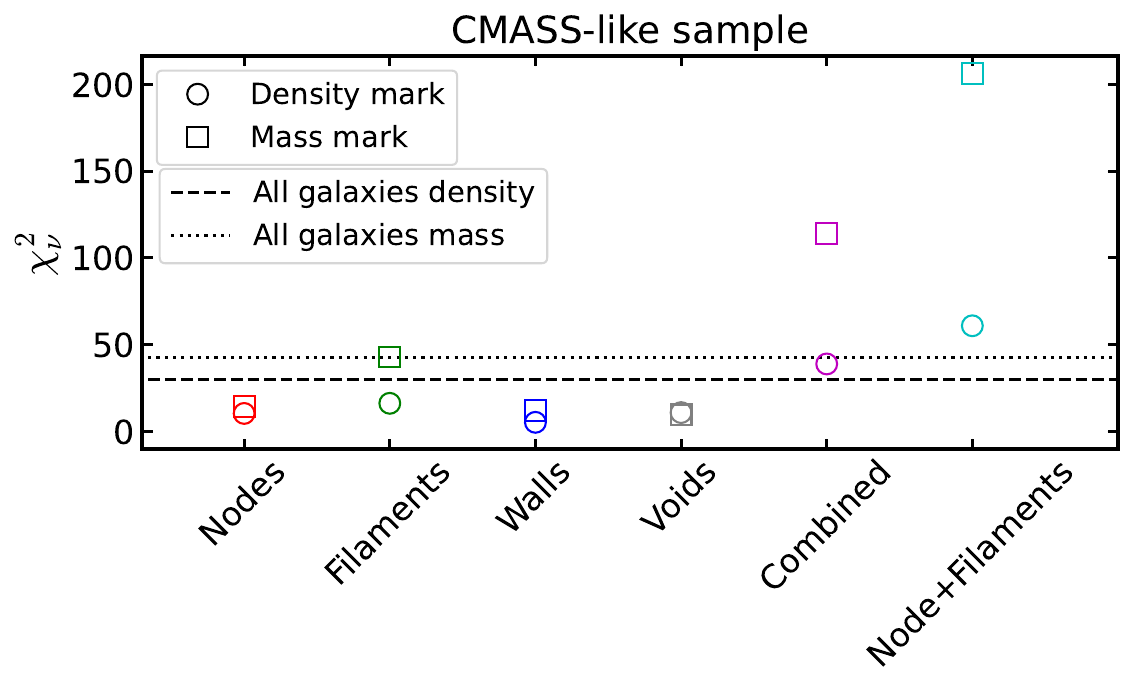}
    &
    \includegraphics[width=0.445\linewidth]{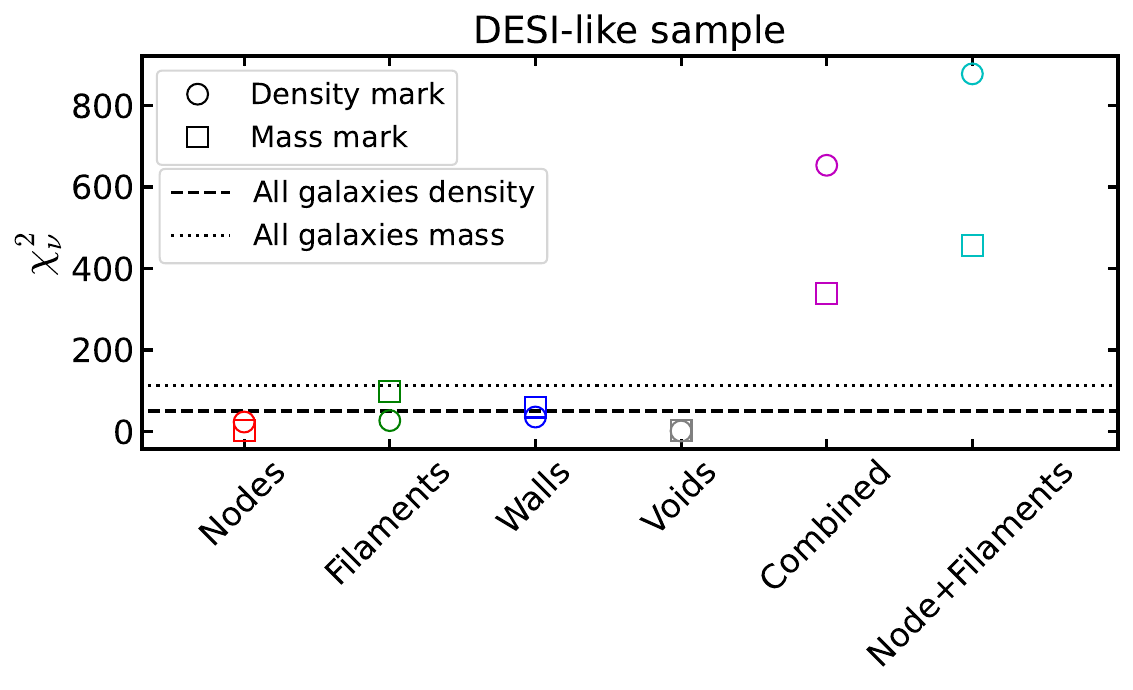}
    \end{tabular}
    \caption{Reduced chi-square statistic $\chi_{\nu}^2$ of $\mathcal{M}$ for different cosmic structures (same colour scheme as Figure \ref{fig:dn_dlogrho}) and some data vector combinations. We compare values for both density (open circles) and mass (open squares) marks and add the $\chi_{\nu}^2$ values of density (dashed line)  and mass-marked (dotted line) correlations using all galaxies. We choose to add cases where we calculate $\chi_{\nu}^2$ using the combination of individual data vectors (magenta): $\mathcal{M}^{\rm node},\ \mathcal{M}^{\rm filament},\ \mathcal{M}^{\rm wall},\ \mathcal{M}^{\rm void}$ and the combination of $\mathcal{M}^{\rm node}$, and $\ \mathcal{M}^{\rm filament}$ (cyan).}
    \label{fig:redchisquare_Nodesfilaments}
\end{figure*}

To further determine whether the individual information provided by individual environments can \newtext{improve the sensitivity to deviations from GR} of already existing test, such as the marked correlation function from \cite{Satpathy2019} and  \cite{Armijo2024a}, we calculated the individual reduced chi-square ($\chi_{\nu}^2$) of the data vectors for individual structures and some potential useful combinations. We also consider the uncertainties related to model galaxies using the HOD prescription, \newtext{distinguishing two separate contributions: the variance produced by using different HOD parameter values (HOD parameter uncertainty) and the noise introduced by the HOD stochasticity at fixed HOD parameters, both explored in previous studies} \citep{Armijo2024b}. These contributions are smaller than 5\% and 1\% respectively, on the scales studied throughout this paper. \newtext{When combining probes, for instance nodes and filaments measurements into a joint data vector $\mathcal{M}^{\rm node+filament}$, the concatenated data vector spans 30 bins and a new $30\times 30$ bins covariance matrix. This joint covariance naturally encodes the statistical dependence between the two auto-correlations, with the off-diagonal blocks capturing the cross-covariance between environments}. In Figure~\ref{fig:redchisquare_Nodesfilaments}, we find the values of $\chi_{\nu}^2$ for both types of marks calculated in all samples. Broadly, the individual components have a similar, but smaller values than the test using all galaxies, being equal only in the case of filaments in particular when using mass type marks. However, these values are improved once the data vectors of all structures are combined when doing the same analysis, by a factor of $2.3$ larger. Following the same pattern, the test clearly over-performs for the particular combination of nodes and filaments cosmic structures, by a factor $4.1$, increasing the amount of information obtained by these data vectors combinations. When considering the same test for a DESI-like sample in the right hand side panel of Figure~\ref{fig:redchisquare_Nodesfilaments} a similar behaviour is found. Nevertheless, as the DESI-like sample, at higher redshift has a higher $n_{\rm gal}$ by around 30\% this is translated as a factor $\times$4 larger in terms of $\chi_{\nu}^2$, which is more evident for $\mathcal{M}$ when combining data vectors, with $\chi_{\nu}^2 = 400$ for all structures combined and $\chi_{\nu}^2 = 800$ for $\mathcal{M}^{\rm node} + \mathcal{M}^{\rm filament}$ case. However, there is a reversed trend, where the density marks have higher $\chi_{\nu}^2$ for combined probes. We attribute this to the mass mark being less effective at higher redshift as modified gravity enhancement is less effective on these halo masses at $z=1.0$.

\section{Conclusions}\label{sec6}

We analyse simulations of the HS $f(R)$ modified gravity model with a fifth-force amplitude of $\left| f_{R0} \right| = 10^{-5}$ compared to a standard GR counterpart. Using the multi-scale morphological filter code \texttt{pycosmommf}, we classify the simulations into distinct cosmic structures, thereby defining the environments in which haloes and galaxies reside. Mock galaxy samples with properties matching those of CMASS and DESI LRGs are then constructed, and marked statistics are applied as environment-dependent tests to distinguish between MG and GR. Therefore, this study is a direct application of the marked correlation functions introduced in \cite{Armijo2018}. Our main findings are summarized as follows:

\begin{itemize}
    \item Cosmic structures provide a meaningful definition of environments in both GR and F5 simulations, as their classification is based on the matter density field, tidal field and its derivatives (via the Hessian).
    \item Galaxies residing in different environments exhibit distinct properties: they become more massive in unscreened regions, such as cosmic voids where the fifth force is active, while galaxies in screened regions remain unaffected.
    \item The marked correlation function applied to individual cosmic structures effectively distinguishes F5 from GR, whether galaxies are marked by local density or by host halo mass.
    \item For filaments and cosmic voids, the marked correlation function (with density and mass marks) yields the highest signal-to-noise ratios, on scales of $1 < r / ({\rm Mpc}\, h^{-1}) < 5$ and $5 < r / ({\rm Mpc}\, h^{-1}) < 40$, respectively.
    \item \newtext{The sensitivity to F5 deviations from GR} increases when their independent marked correlation function measurements are combined, raising the reduced chi-square $\chi_{\nu}^2$ by a factor of 2.3 relative to the case using all galaxies.
    \item Combining the marked correlation functions of nodes and filaments further improves the fit, exceeding the case with all galaxies by more than a factor of four. This reduced $\chi^2$ highlights the untapped potential of filamentary structures as promising regions of interest in modified gravity simulations.
    \item Forecasts based on current catalogues, such as the DESI LRG sample, show that \newtext{the sensitivity} of the of the marked correlation function can increase by a factor of four compared to the CMASS sample.
\end{itemize}

The marked correlation function results for independent cosmic structures, as well as their combined data vectors, demonstrate the advantages of using filaments as enhanced environments in tests of modified gravity. While cosmic voids dominate the linear regime and therefore hold promise for future constraints within linear theory, the the strongest sensitivity to F5 deviations from GR is expected to come from the quasi-linear and non-linear information provided by nodes (e.g., galaxy clusters and large mass haloes) and filaments.

We attribute the improvements from including filaments to both the larger number of galaxies residing in filaments and the distinct properties of filaments in F5 relative to GR simulations. The marked correlation function, specifically designed to capture such differences, enhances the constraining power when evaluated in scales where the fifth force is stronger. To estimate the information content via the expected S/N ratio (analogue to Fisher analysis) we estimate the covariance matrix using Jackknife resampling, \newtext{complemented by the impact of the HOD stochastic contribution at fixed HOD parameters on the mark distribution.} We acknowledge that this approach is simpler than alternatives such as covariance estimation from independent realizations (Poisson) for the full simulations; however, the possible resulting bias primarily affects smaller scales ($r < 1\ {\rm Mpc}\, h^{-1}$) than the ones used on this paper, as shown in uncertainties of clustering studies \citep{Norberg2009}. We provide the covariance matrices used in the analysis, including the corresponding correlation coefficients, in the Appendix \ref{apx:Covariance}.

For the construction of mock samples resembling observations, we follow the methodology of \cite{Cautun2018}, which calibrates the HOD parameters applied to the F5 simulation to reproduce both the number density and clustering of the corresponding GR sample. We adopt number densities consistent with recent LRG samples, such as CMASS and DESI-LRG. A limitation of this approach is that it relies only on simulation outputs at redshift $z=0.0$ and $z=1.0$, which simplifies the analysis. \newtext{The snapshots adopted here do not precisely match the effective redshift of the CMASS and DESI LRG samples, and these mock catalogues should be understood as referring to number-density-matched analogues rather than exact survey mocks. While this redshift mismatch introduces some simplification, it does not affect our main conclusion regarding the relative constraining power of filaments in the marked correlation function as the environmental sensitivity of the signal is expected to persist across this redshift range}. In a realistic application where this test is used to constrain the amplitude of the fifth force, the inclusion of filaments is still expected to yield significant improvements.

In synthesis, our method provides a clear way to incorporate the information from filaments, nodes, walls an voids into modified gravity studies, offering a pathway to improved constraints with current and future surveys. This approach is complementary, yet particularly relevant for forecasts with stage-IV surveys, such as the Euclid mission \citep{Koyama2024} and Rubin–LSST \citep{Davies2024}. \newtext{A natural extension of this work is to explore the scaling of the environmental marked correlation function signal with $f_{R0}$, simultaneously varying cosmological parameters, which demands emulator-based simulations suites.} By explicitly accounting for the role of filamentary environments, our approach enhances the sensitivity of marked statistics to fifth-force effects beyond what can be achieved with traditional two-point statistics alone. We therefore expect that the integration of filament-based marked correlation functions into future analyses will become a powerful tool for testing gravity on cosmological scales.

\section*{Acknowledgements}
The authors would like to thank Daniela Galarraga-Espinoza for useful conversations and to Baojiu Li for providing part of the simulation data and helping to improve the quality of the manuscript. JA is supported by JSPS KAKENHI Grant JP23K19064. This work was supported by the World Premier International Research Center Initiative (WPI), MEXT, Japan. LDC acknowledge financial support from the ILANCE initiative. This work used simulations run at the DiRAC@Durham facility managed by the Institute for Computational Cosmology on behalf of the STFC DiRAC HPC Facility (www.dirac.ac.uk). The equipment was funded by BEIS capital funding via STFC capital grants ST/K00042X/1, ST/P002293/1, ST/R002371/1 and ST/S002502/1, Durham University and STFC operations grant ST/R000832/1. DiRAC is part of the National e-Infrastructure.

\section*{Data Availability}
The simulations used on this study are available and accessible on reasonable request. The data products of this work can be shared upon request to the corresponding author.



\bibliographystyle{mnras}
\bibliography{main} 




\appendix

\section{Covariance matrix} \label{apx:Covariance}

We calculate the covariance matrix of our measurements of $\mathcal{M}(r)$ using the Jackknife resampling, also known as the 'leave-one-out' method. By dividing the simulation box in $N = 512$ sub samples with equal volume, we calculate the marked correlation function $\mathcal{M}_i$ of a volume equals to the box volume minus the subvolume, in other words we omit the $i$-subvolume when calculating our data vector. In this way the covariance matrix can be expressed as:
\begin{equation}
    C_{ij} = \frac{N}{N-1}\sum_k^N (\mathcal{M}_{i}^{k} - \bar{\mathcal{M}_i})(\mathcal{M}_j^k - \bar{\mathcal{M}_j})
\end{equation}

\begin{figure}
    \centering
    \includegraphics[width=0.99\linewidth]{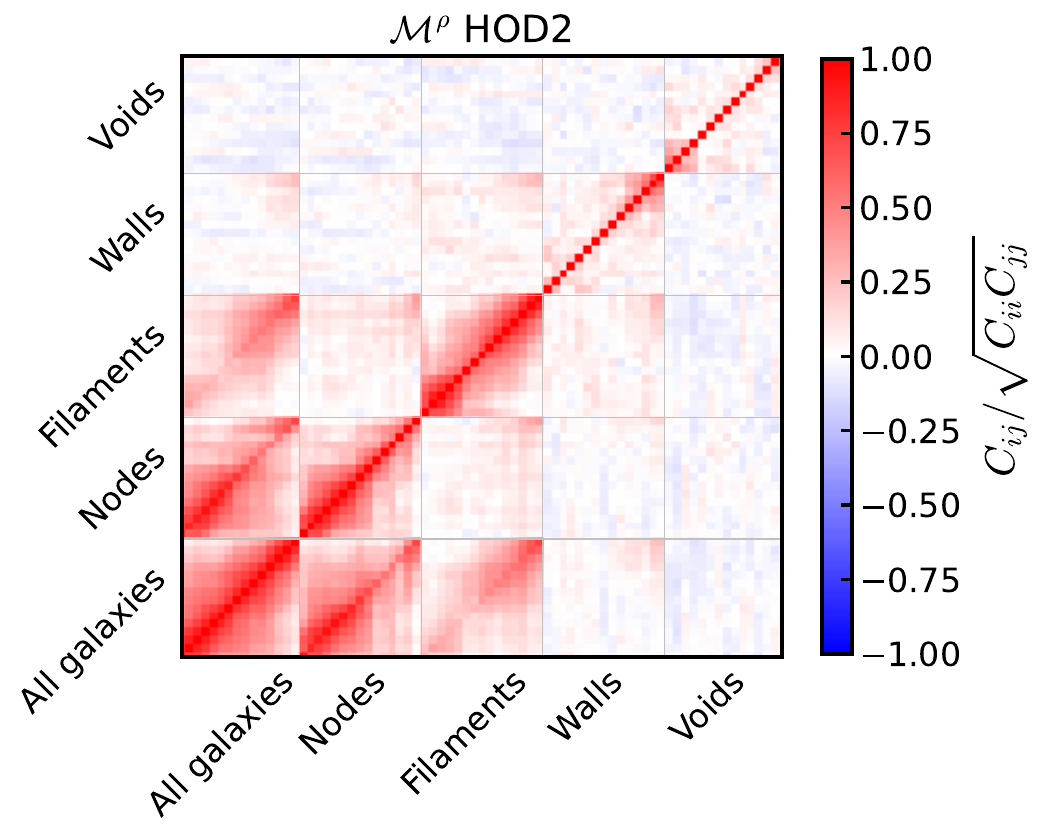}\\
    \includegraphics[width=0.99\linewidth]{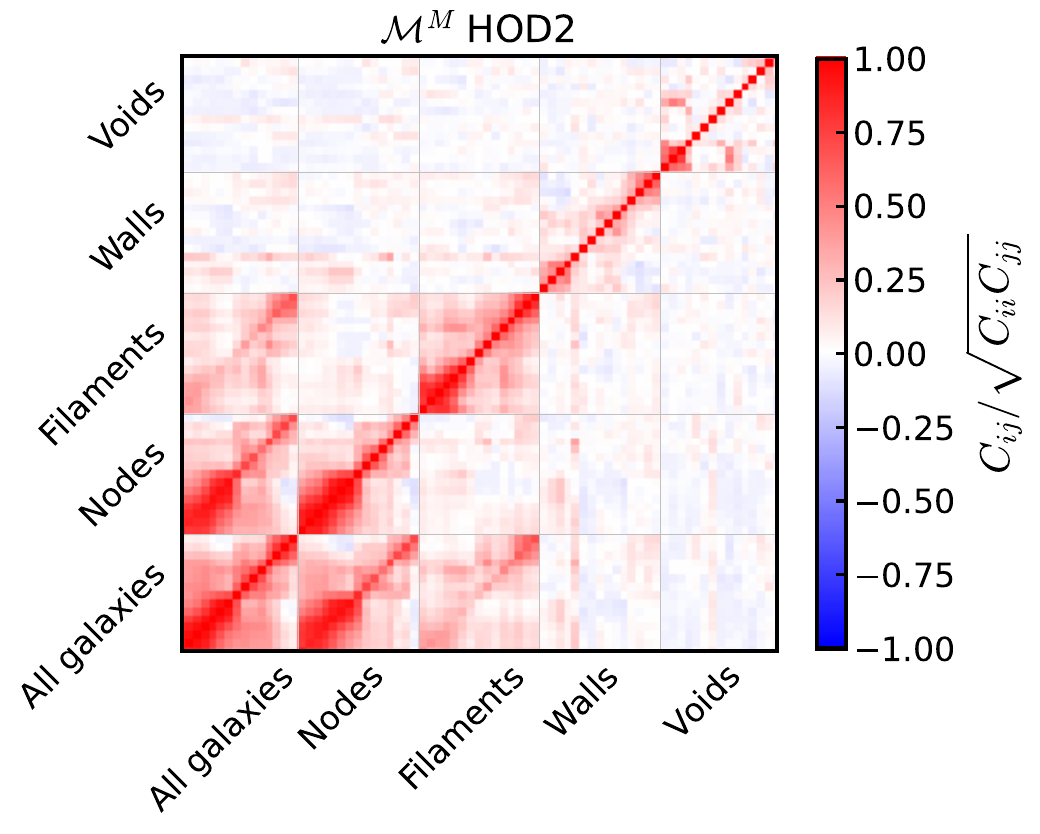}
    \caption{The correlation coefficients of the covariance matrix of the density-marked (top) $\mathcal{M}^{\rho}$ and mass-marked $\mathcal{M}^{M}$ (bottom) correlation functions. We include the measurements using all galaxies (from the HOD2 sample) and for results for individual cosmic structures.}
    \label{fig:covariance_marked_densityMass}
\end{figure}

Where $N$ is the number of subsamples and $\bar{\mathcal{M}}=\frac{1}{N}\sum_k \mathcal{M}^k$ the mean of the ensemble. The extra factor $N$ appearing in the fraction numerator accounts for the lack of independency of the samples. We show the covariance matrix of $\mathcal{M}$ in Figure \ref{fig:covariance_marked_densityMass} where the correlation coefficients are showed for the marked correlation functions used throughout this paper. Similar results are obtained for both samples HOD1 and HOD2. Even though a high degree of correlation is found when using $\mathcal{M}$ all galaxies this decreases for some of the structures, particularly walls and voids. \newtext{The correlation coefficients show that the environments considered provide some statistical information, justifying the gain in S/N achieved when combining data vectors.}

\bsp	
\label{lastpage}
\end{document}